\documentclass[conference]{IEEEtran}
\IEEEoverridecommandlockouts
\usepackage{cite}
\usepackage{amsmath,amssymb,amsfonts}
\usepackage{algorithmic}
\usepackage{textcomp}
\usepackage{xcolor}
\usepackage{graphicx}
\usepackage{epstopdf}
\usepackage{subfig}
\usepackage{color}

\newtheorem{lemma}{\textbf{Observation}}

\def\BibTeX{{\rm B\kern-.05em{\sc i\kern-.025em b}\kern-.08em
		T\kern-.1667em\lower.7ex\hbox{E}\kern-.125emX}}
\begin{document}
	
	\title{A Deep Dive into Blockchain Selfish Mining}
	
	\author{\IEEEauthorblockN{Qianlan Bai$^+$, Xinyan Zhou$^+$, Xing Wang$^{\dag}$, Yuedong Xu$^+$, Xin Wang$^{\dag}$, Qingsheng Kong$^+$}
		\IEEEauthorblockA{$^+$ School of Information Science and Engineering, Fudan University} 
		$^{\dag}$ School of Computer Science and Technology, Fudan University}

	\maketitle
	
	\begin{abstract}
		This paper studies a fundamental problem regarding the security of blockchain on how the existence of multiple misbehaving pools influences the profitability of selfish mining. Each selfish miner maintains a private chain and makes it public opportunistically for the purpose of acquiring more rewards incommensurate to his Hashrate. We establish a novel Markov chain model to characterize all the state transitions of public and private chains. The minimum requirement of Hashrate together with the minimum delay of being profitable is derived in close-form. The former reduces to 21.48\% with the symmetric selfish miners, while their competition with asymmetric Hashrates puts forward a higher requirement of the profitable threshold. The profitable delay increases with the decrease of the Hashrate of selfish miners, making the mining pools more cautious on performing selfish mining.
	\end{abstract}
	
	
	\section{Introduction}
	Bitcoin has gained tremendous concerns as the first fully decentralized cryptocurrency since its advent in 2008. 
	All historical transactions between Bitcoin clients are recorded in a global and public data structure known as the
	\emph{blockchain}. The security of the blockchain is established by a chain of cryptographic Hash puzzles, addressed by 
	a large-scale network of pseudonymous participants called \emph{miners} \cite{ref:whitepaper}. 
	Solving a Hash puzzle is deemed as a way to generate Proof-of-Work (PoW) of reaching global consensus. 
	The PoW of Bitcoin demands intensive computations, thus consuming a lot of energy. 
	Each miner competes for this ``game'', and is rewarded by cryptocurrencies (i.e. bitcoins) if he is the first acknowledged 
	miner to find a valid block. When the population of miners is large, the aggregate Hash power is sufficiently high such 
	that a malicious miner can hardly accumulate enough Hash power to perform Sybil attacks. 
	The PoW consensus of Bitcoin has been employed in almost $90$\% of public blockchains, serving as the cornerstone of 
	current cryptocurrencies. 
	
	The security of PoW is challenged by the trend of centralization of Hash power. Mining a Bitcoin block is random and it needs 
	more than 10 years on average with a latest-generation ASIC chip. 
	Therefore, blockchain miners operate strategically to form pools that have a much larger chance of solving puzzles in each round. By splitting the mining reward appropriately, they acquire a stable income rate. 
	As a side effect, a small number of mining pools occupy a vast majority of global Hash power, placing blockchain 
	systems at the risk of being overthrown by a gigantic pool or colluding pools. 
	The conventional wisdom believes that PoW is secure as long as no miner controls 51\% of total Hash power. 
	However, a miner can choose a \emph{selfish mining} scheme instead of conforming to the standard Bitcoin protocol.
	Eyal and Sirer pointed out that the selfish mining is profitable (i.e. more rewards than the honest mining) if the Hash power
	of a miner is larger than 25\% \cite{ref:majority}. A more intelligent selfish miner using Markov Decision Process (MDP) can lower down this 
	threshold to around 23.21\% \cite{ref:MDP}. Note that the both studies assume the existence of a single selfish miner while 
	multiple (colluding) pools might be close to this profitable threshold. 
	
	In this paper, we study a fundamental question regarding the blockchain security: 
	\emph{Will selfish mining become more easily profitable when there exist more than one selfish miners, and 
		how many rounds should a selfish miner wait until being profitable?} 
	The former subquestion aims to unravel whether each selfish miners needs a smaller threshold of Hashrate to 
	gain more rewards than mining honestly. The latter pays attention to the transient behavior in the process of selfish mining 
	that takes into account the mining difficulty adjustment. The transient analysis is also crucial for a selfish miner 
	is inclined to waiting for a long period to gain more rewards, especially when the global Hashrate increases rapidly. 
	We establish the selfish mining model for an honest pool that represents all honest miners, and two 
	selfish mining pools who are not aware of each other's misbehaving role. 
	By dissecting all the possible events that trigger the change of private and public chains, 
	we formulate a set of Markov chains to capture all the state transitions. 
	In contrast to a very recent experimental study\cite{ref:ruanna} that analyzes the profitable threshold of 
	selfish mining with two miners, our work presents a mathematical model that yields close-form expression of such a threshold. 
	In the transient analysis, the selfish mining is found of wasting computing power and thus is definitely unworthy 
	without the subsequent difficulty adjustment of puzzle-solving.  
	
	The major contributions and observations are summarized as below. 
	
	\begin{itemize}
		
		\item We establish a set of Markov chain models to characterize the state transition of public 
		and private chains in selfish mining and compute the steady state distributions.

		\item The minimum threshold of Hashrate is symmetric around 21.48\% if two selfish miners are both profitable. 
		While the profitable selfish mining becomes more difficult when one of the selfish miner increases his Hashrate,  
		arousing a more furious competition.  
		
		\item  The selfish mining is profitable after 51 rounds of difficulty adjustment (i.e. 714 days in Bitcoin) if the Hashrates of selfish miners are both 22\% (slightly higher than the profitably threshold). This delay decreases to 
		5 rounds (i.e. 70 days in Bitcoin) as their Hashrates accrues to 33\%, which is still very long. 
		
	\end{itemize}
	
	\section{System Model}
	
	In this section, we describe the basic model of blockchain mining in the presence of 
	two adversarial pools. 
	
	\subsection{System Description}
	
	Consider a blockchain mining system with two misbehaving mining pools \emph{Alice} and \emph{Bob}, as well as
	an honest mining pool, $\emph{Henry}$\footnote{Multiple honest miners can be boiled down to a single miner for the sake of their linear additivity of Hashrates.}. They compete to solve cryptographic puzzles to mine a valid block for the purpose of acquiring bitcoin-like rewards. The proof-of-work (PoW) consensus is adopted and the mining of blocks is stateless: 
	the probability of discovering a block by 
	a miner is proportional to his current Hashrate, but inversely proportional to the current aggregate Hashrate of the entire blockchain network. The blockchain system dynamically adjusts the difficulty of cryptographic puzzles 
	such that new blocks are generated at a fixed average rate(e.g. one block per 10 minutes on average in Bitcoin).
	We define a ``round" as the time to process one attack.
	The miners maintain a globally-agreed ordered set of transactions via the adoption and the mining 
	on the longest chain. The revenue of a miner
	is the expected fraction of blocks mined by him out of all the blocks in the longest chain
	
	For the simplicity, we make the following assumptions that are consistent with the literature \cite{ref:majority}.
	\begin{itemize}
		\item The total Hashrate of the blockchain system is normalized as a unit. Then, the Hashrate of a mining pool is represented as a fraction of the total.
		\item The block discovery time by a mining pool is exponentially distributed when his Hashrate is large.
		\item The reward of each valid block is normalized as one cryptographic coin. 
	\end{itemize}
	
	Denote by ${\alpha}_{1}$, ${\alpha}_{2}$ and ${\alpha}_{h}$ the Hashrates of Alice, Bob and Henry respectively, i.e. \begin{math}\alpha_1+\alpha_2+\alpha_h=1\end{math}. 
	Denote by $\gamma_1$ (resp. $\gamma_2$) the probability that honest miners mine after Alice's (resp. Bob's) released chain in the tie-breaking between Alice (resp. Bob) and Hence. Denote by $\theta_1$ and $\theta_2$ the probabilities that honest moners choose to mine after Alice's and Bob's chains in the three-party tie-breaking, respectively. When the blockchain system creates a new block, it is mined by pool $i$ with the probability $\alpha_i$ , $\forall i\in\{1,2,h\}$, owing to the memorylessness of exponentially distributed mining intervals.

	Alice (resp. Bob) may release her blocks strategically by forcing Henry into wasting his computations. When Alice and Bob are both selfish miners, the interaction between two private chains becomes more complicated because none of them know other's behaviour. In what follows, we capture all the different states that each miner may encounter.
	


	\subsection{Selfish Mining Mode}
	
	Alice maintains a private chain, so does Bob, while Henry operates on the public chain. 
	Alice and Bob are not aware of each other's role. 
	We suppose that all the miners work on the same public chain in the beginning where the 
	starting point is expressed as ``0''. The length of the private 
	chain is kept as a private information by Alice and Bob, and the length of the public chain 
	is observed by all of them. We consider the selfish mining method proposed by \cite{ref:majority}, and our analytical approach 
	can be generalized to a variety of other methods. 
	
	The \emph{mining procedure} consists of two cases as follows.
	\begin{itemize}
		
		\item \emph{(Public-chain mining case)} Henry always mines after the public chain. Alice or Bob 
		also mines on the public chain if it is longer than his private chain. 
		
		\item \emph{(Private-chain mining case)} Alice (resp. Bob) continues to mine on her (resp. his) private chain if she (resp. he) discovers a new block and the private chain is now longer than the public chain. 
		
	\end{itemize}
	
	The \emph{release procedure} is more complicated than the mining procedure. Henry broadcasts his mined block 
	as soon as it is discovered, while Alice and Bob will decide whether to release their mined blocks depending on the length of the public chain. 
	
	\begin{itemize}
		
		\item \emph{(Forfeit case)} Alice (resp. Bob) abandons her (resp. his) private chain and conforms to 
		mining after the public chain if the latter is longer. Henry also abandons his public chain if Alice or Bob publishes a 
		longer chain. 
		
		\item \emph{(Risk-avoiding release case)} Alice (resp. Bob) releases her (resp. his) privately mined blocks to the public because of the fear of loss if the 
		new block is mined by the others and the leading advantage of her private chain is no more than two blocks. 
		
		\item \emph{(Chain reaction case)} When Alice (resp. Bob) releases her (resp. his) blocks to the public chain and updates its 
		length, the release of Bob's (resp. Alice's) private blocks is triggered immediately. 
		
	\end{itemize}
	
	The chain reaction case is the combination of the forfeit and the risk-avoiding cases, whereas the existence of chain reaction complicates evolution of the public chain. 
	Suppose that Alice publishes her private blocks to obsolete the current public chain. 
	After the construction of new public chain, Bob may release his private chain to forfeit it immediately.

	\subsection{Release procedure and tie-breaking Logics}
	
	The consensus on the public chain requires that it is the longest. A crucial question is how the public chain 
	evolves when it is of the same length as Alice or Bob. In general, each miner works on his own chain, and the release behavior 
	of Alice and Bob is triggered when Henry mines a new block. We hereby illustrate the evolution of private and public chains 
	where $A_k$, $B_k$, and $H_k$ denote that the $k^{th}$ blocks belong to Alice, Bob and Henry respectively. 
	The blocks of private chains are in grey and those of public chains are in white. 
	
\begin{figure}[!ht]
		\centering
		\includegraphics[width=3in]{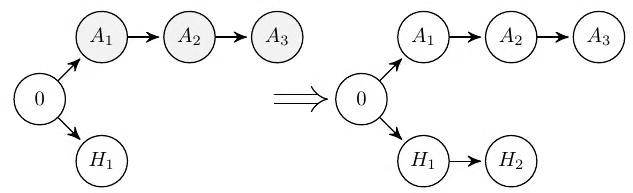}
		\caption{Alice's risk-avoiding release and Henry's abandonment.}
		\label{fig:forfeit}
		\vspace{-0.2in}
	\end{figure}

	\noindent\textbf{Risk-avoiding release case} 
	We show the risk-avoiding release of Alice's private chain in Figure \ref{fig:forfeit}. 
	Alice is only one block ahead of Henry after the latter mines a new block for the public chain. 
	Because Alice fears of losing the competition, she publishes her private blocks, obseleting Henry's public chain, so that 
	both Alice and Henry mine on the new longest chain afterwards. 
	
\begin{figure}[!ht]
		\vspace{-0.1in}
		\centering
		\includegraphics[width=2.5in,height=1.35in]{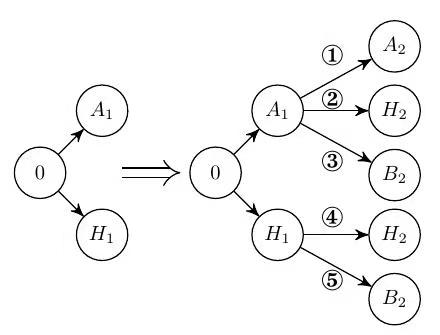}
		\caption{Tie-breaking case of two public chains.}
		\label{fig:tie_two}
		\vspace{-0.1in}
	\end{figure}

	\noindent\textbf{Tie-breaking resolvings.} If Alice's private chain is only one block ahead of Henry's, Henry may catch up with her. When it happens, Alice publishes her private blocks immediately to compete with Henry. Thus, two public chains of the same length exist in Figure \ref{fig:tie_two}. Since only one public chains prevails, a tie-breaking rule needs to be taken into account. The first case is that the public chains of Alice and Henry have the same length, and Bob's private chain is either 0 or very long. Hence, we only need to resolve the tie between Alice and Henry. All the miners are possible to mine after block $A_1$, while Bob and Henry may mine after $H_1$. There are five possibilities of extending the longest public chain, and the shorter one will be obsoleted. We omit the tie-breaking between Bob and Henry because this can be analyzed in the same way. 
	
\begin{figure}[!ht]
		\vspace{-0.2in}
		\centering
		\includegraphics[width=3in,height=1.35in]{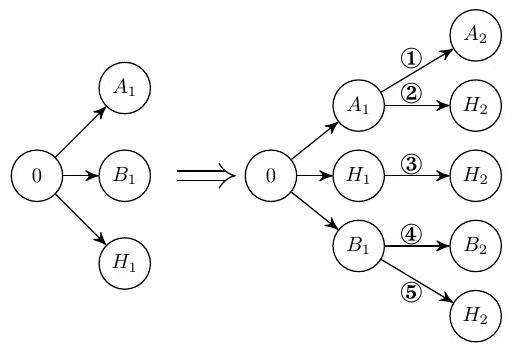}
		\caption{Tie-breaking case of three public chains.}
		\label{fig:tie_three}
	\end{figure}

	For the situation that each of Alice and Bob hides one private block, they will publish their private chains instantly after Henry 
	finds a new block. As shown in Figure \ref{fig:tie_three}, there exists three competing public chains. 
	Alice will mine after $A_1$ and Bob will mine after $B_1$ for sure; Henry is not aware of which chain is maliciously forked so 
	that he may mine on each public chains. There are also five possible situations. 
	The risk-avoiding release, together with two tie-breaking solutions, constitutes all the dynamics of private and public chains. 
	\vspace{-0.1in}
	
	\begin{figure}[!ht]
		\centering
		\includegraphics[width=3in]{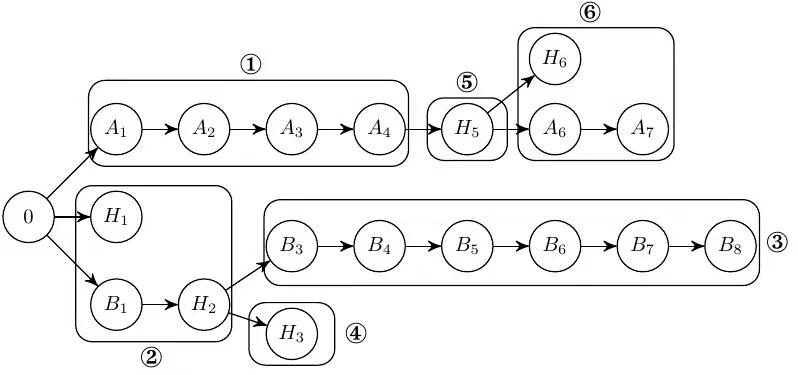}
		\caption{Chain reaction case.}
		\label{fig:chain_reaction}
		\vspace{-0.2in}
	\end{figure}
	
	\noindent\textbf{Chain reaction release.} 
	We next introduce the chain reaction release that complicates the evolution of the private and public chains. 
	Note that the chain reaction release consists of a sequence of risk-avoiding releases and tie-breaking resolvings. 
	Figure \ref{fig:chain_reaction} illustrates an example on how the chain reaction 
	phenomenon is triggered. At stage 1, Alice's private chain contains four blocks while the lengths of Bob's private chain 
	and Henry's public chain are 0. After a tie-breaking resolving at stage 2, the longer public chain contains two blocks 
	$B_1$ and $H_2$, and the shorter is orphaned. Bob construct a new private chain starting from $B_3$ to $B_8$, 
	while Henry continues to mine one block after $H_2$ at stage 4. 
	From Alice's perspective, her private chain is merely one block ahead of the public chain. She releases her private blocks 
	in order to avoid the risk of losing the race with Henry. The new public chain now starts from block $A_4$. 
	Next, stage 5 and 6 constitute a new round of tie-breaking resolving between Alice and Henry, extending the public chain 
	to block $A_7$. However, the release of $A_7$ triggers Bob to release all of his private blocks starting from $B_3$ to 
	$B_8$. When retrospecting all the mining stages, we observe that the winning branch switches back and forth, 
	making the analysis of selfish mining extremely complicated. 
	
	\section{Finite State Machine}
	In this section, we construct the state machine of blockchain selfish mining and 
	present the steady-state and transient analysis of the profitable threshold.
	
	\subsection{Steady-state Analysis}
	\begin{figure}[h]
		\vspace{-0.2in}
		\centerline{\includegraphics[height=0.2\textheight,width=0.4\textwidth]{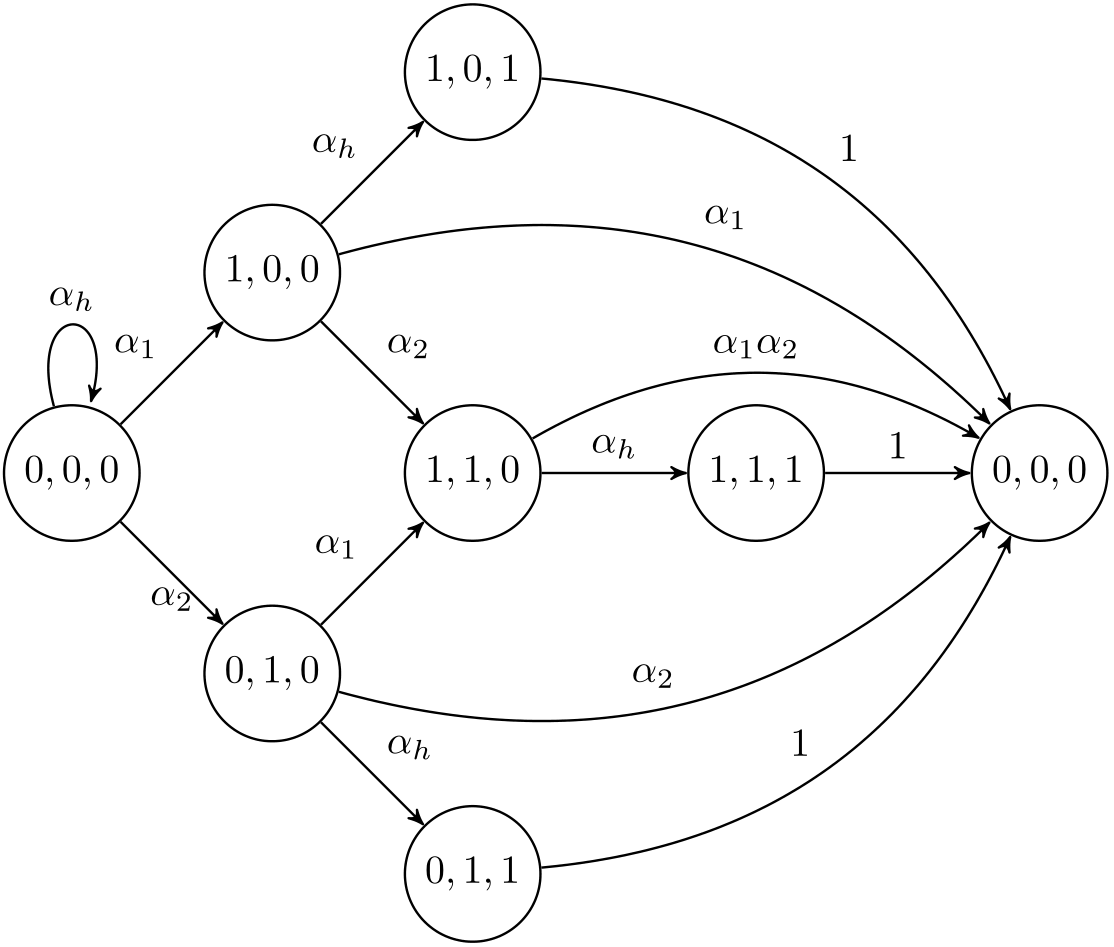}}
		\caption{State machine with $N{=}2$.}
		\label{fig}
		\vspace{-0.2in}
	\end{figure}
	
	We hereby formulate a finite state machine to characterize the evolution of private and public chains. 
	Figure \ref{fig} illustrates the state machine when the maximum length of private chain is two (i.e. $N{=}2$). 
	We define the state as a three-tuple consisting of the lengths of Alice, Bob and Henry. 
	The arrows indicate the corresponding state transitions and the associated values represent the 
	transition probabilities. For instance, all the transitions to $(0,0,0)$ mean that 
	the forked chains boil down to the unanimous public chain and a new round of selfish mining starts. 
	Denote by $P_{ijk}$ the steady state distribution of $(i,j,k)$. Denote by $R_1$ (resp. $R_2$, $R_h$) 
	the average number of valid blocks mined by Alice (resp. Bob, Henry). 
	Using the standard approach, we obtain $R_{i}$ as follows \cite{ref:state}.

	\vspace{-0.5cm}
	\begin{eqnarray}
	P_{000}^{{-}1}=
	1{+}\alpha_1{+}\alpha_2{+}\alpha_1\alpha_h{+}2\alpha_1\alpha_2{+}\alpha_2\alpha_h {+}2\alpha_1\alpha_2\alpha_h;
	\end{eqnarray}
	\vspace{-0.5cm}
	\begin{eqnarray}
	\frac{R_1}{P_{000}}=
	&&\!\!\!\!\!\!\!\!\!\!\!2\alpha_1^2\left(1+\alpha_h \right)+\left(\alpha_2+\alpha_h \right)\alpha_1\alpha_h\gamma_1+\alpha_1\alpha_2\alpha_h \nonumber\\
	&&\!\!\!\!\!\!\!\!\!\!\!+4\alpha_1^2\alpha_2\left(1+\alpha_h \right)+2\alpha_1\alpha_2\alpha_h^2\theta_{1}; \\
	\frac{R_2}{P_{000}}=
	&&\!\!\!\!\!\!\!\!\!\!\!2\alpha_2^2\left(1+\alpha_h \right)+\left(\alpha_1+\alpha_h \right)\alpha_h\alpha_2\gamma_2+\alpha_1\alpha_2\alpha_h \nonumber\\
	&&\!\!\!\!\!\!\!\!\!\!\!+4\alpha_2^2\alpha_1\left(1+\alpha_h \right)+2\alpha_1\alpha_2\alpha_h^2\theta_{2};\\
	\frac{R_h}{P_{000}}=
	&&\!\!\!\!\!\!\!\!\!\!\!\alpha_1\alpha_h^2\left(2-\gamma_1 \right)+2\alpha_1\alpha_2\alpha_h^2\left(2{-}\theta_{1}-\theta_{2} \right)+\alpha_h\nonumber\\
	&&\!\!\!\!\!\!\!\!\!\!\!+\alpha_2\alpha_h^2\left(2-\gamma_2 \right)+\alpha_1\alpha_2\alpha_h(2-\gamma_1-\gamma_2).
	\end{eqnarray}
	When $N$ is large (e.g. three or four), the finite state machine 
	becomes more complicated. Due to limite space, we leave the detailed analysis in the technical report \cite{ref:appendix}, 
	while only presenting the close-form results with $N{=}4$. 
	
	\vspace{-0.5cm}
	\begin{eqnarray}
	\frac{R_1}{P_{000}}=
	&&\!\!\!\!\!\!\!\!\!\!\!4\alpha_1^4\left(1+\alpha_h\right)+3\alpha_1^3\alpha_h^2+16\alpha_1^4\alpha_2+4\alpha_1^2\alpha_h\nonumber\\
	&&\!\!\!\!\!\!\!\!\!\!\!+40\alpha_1^4\alpha_2^2\left(1+2\alpha_2\right)+\alpha_1\alpha_2\alpha_h\left(1+\gamma_1+2\theta_{1}\alpha_h\right)\nonumber\\
	&&\!\!\!\!\!\!\!\!\!\!\!+10\alpha_1^2\alpha_2\alpha_h+20\alpha_1^3\alpha_2\alpha_h\left(3\alpha_2+\alpha_1\right)+15\alpha_1^3\alpha_2\alpha_h^2\nonumber\\
	&&\!\!\!\!\!\!\!\!\!\!\!+4\alpha_1^4\alpha_2^2\alpha_h\left(1+\alpha_h\right)+4\alpha_1^4\alpha_2^3\alpha_h^2\left(\beta_1+20\right)\nonumber\\
	&&\!\!\!\!\!\!\!\!\!\!\!+5\alpha_1^5\alpha_2^3\alpha_h+4\alpha_1^4\alpha_2^3\alpha_h\left(\alpha_2+21\right)+3\alpha_1^3\alpha_2^4\alpha_h^2\beta_1\nonumber\\
	&&\!\!\!\!\!\!\!\!\!\!\!+\alpha_1\alpha_h^2\gamma_1+12\alpha_1^2\alpha_2^2\alpha_h^2\beta_1+\alpha_1^2\alpha_2^2\alpha_h^3\beta_1\left(3\alpha_1+2\alpha_2\right)\nonumber\\
	&&\!\!\!\!\!\!\!\!\!\!\!+6\alpha_1^3\alpha_2^3\alpha_h^2\left(10\alpha_h\beta_1+1\right);\\
	\frac{R_2}{P_{000}}=
	&&\!\!\!\!\!\!\!\!\!\!\!4\alpha_2^4\left(1+\alpha_h\right)+3\alpha_2^3\alpha_h^2+16\alpha_1\alpha_2^4+4\alpha_2^2\alpha_h\nonumber\\
	&&\!\!\!\!\!\!\!\!\!\!\!+40\alpha_1^2\alpha_2^4\left(1+2\alpha_1\right)+\alpha_1\alpha_2\alpha_h\left(1+\gamma_2+2\theta_{2}\alpha_h\right)\nonumber\\
	&&\!\!\!\!\!\!\!\!\!\!\!+10\alpha_1\alpha_2^2\alpha_h+20\alpha_1\alpha_2^3\alpha_h\left(3\alpha_1+\alpha_2\right)+15\alpha_1\alpha_2^3\alpha_h^2\nonumber\\
	&&\!\!\!\!\!\!\!\!\!\!\!+4\alpha_1^2\alpha_2^4\alpha_h\left(1+\alpha_h\right)+4\alpha_1^3\alpha_2^4\alpha_h^2\left(\beta_2+20\right)\nonumber\\
	&&\!\!\!\!\!\!\!\!\!\!\!+5\alpha_1^3\alpha_2^5\alpha_h+4\alpha_1^3\alpha_2^4\alpha_h\left(\alpha_1+21\right)+3\alpha_1^4\alpha_2^3\alpha_h^2\beta_2\nonumber\\
	&&\!\!\!\!\!\!\!\!\!\!\!+\alpha_2\alpha_h^2\gamma_2+12\alpha_1^2\alpha_2^2\alpha_h^2\beta_2+\alpha_1^2\alpha_2^2\alpha_h^3\beta_2\left(2\alpha_1+3\alpha_2\right)\nonumber\\
	&&\!\!\!\!\!\!\!\!\!\!\!+6\alpha_1^3\alpha_2^3\alpha_h^2\left(10\alpha_h\beta_2+1\right);
	\end{eqnarray}
	\vspace{-0.7cm}
	\begin{eqnarray}
	\frac{R_h}{P_{000}}=
	&&\!\!\!\!\!\!\!\!\!\!\!\alpha_1\alpha_h^2\left(2-\gamma_1 \right)+\alpha_2\alpha_h^2\left(2-\gamma_2 \right)+\alpha_1^2\alpha_2^3\alpha_h^3\left(2\beta_1\!\!+\!\!\beta_2 \right)\nonumber\\
	&&\!\!\!\!\!\!\!\!\!\!\!+2\alpha_1\alpha_2\alpha_h^2\left(2-\theta_{1}-\theta_{2} \right)+\alpha_1^2\alpha_2^2\alpha_h^2\left(6+4\alpha_1\alpha_2\right)\nonumber\\
	&&\!\!\!\!\!\!\!\!\!\!\!+\alpha_1^3\alpha_2^2\alpha_h^3\left(\beta_1+2\beta_2 \right)+\alpha_1\alpha_2\alpha_h\left(2-\gamma_1-\gamma_2 \right)\nonumber\\
	&&\!\!\!\!\!\!\!\!\!\!\!+\alpha_1^3\alpha_2^3\alpha_h\left(\alpha_1+\alpha_2\right)+\alpha_1^3\alpha_2^4\alpha_h^2\left(2\beta_1+\beta_2 \right)+\alpha_h\nonumber\\
	&&\!\!\!\!\!\!\!\!\!\!\!+\alpha_1^4\alpha_2^3\alpha_h^2\left(\beta_1\!\!+\!\!2\beta_2 \right)\!\!+\!\!20\alpha_1^3\alpha_2^3\alpha_h^3\!\!+\!\!2\alpha_1^4\alpha_2^4\alpha_h\\
	\vspace{0.2cm}
	&&\!\!\!\!\!\!\!\!\!\!\!\beta_1=\gamma_1/(\gamma_1+\gamma_2)\quad\beta_2=\gamma_2/(\gamma_1+\gamma_2).
	\end{eqnarray}
	\vspace{-0.5cm}
	
	
	Note that the cases with $N{>}4$ are not considered in the modeling. Apart from their complexity, a large $N$ may 
	cause a lot of consecutively orphaned blocks so that the selfish mining can be easily detected. Later on, 
	our simulation confirms convergence of profitable threshold at $N{=}4$, i.e. the difference between 
	$N{=}4$ and a large enough $N$ is very small.

	\subsection{Transient State Analysis}
	According to the data from \cite{ref:data}, the Hashrate of the Bitcoin system grows exponentially. It is necessary to study the transient behavior of an attack. We model the action during one difficulty adjustment period and explore the relationship between the number of periods and the attackers' Hashrate.
	
	For a better description, we define the concept of \emph{absolute revenue} and \emph{relative revenue}. First, Alice's, Bob's and Henry's \emph{relative revenue} are the proportion of their revenue to total revenue, which are
	\vspace{-0.4cm}
	\begin{eqnarray}
	&&R_A={R_1}/\left(R_1+R_2+R_h\right)\\
	\label{eq:RA}
	&&R_B={R_2}/\left(R_1+R_2+R_h\right)\\
	&&R_H={R_h}/\left(R_1+R_2+R_h\right)
	\end{eqnarray}
	
	Since we ignore the influence of transaction fee and other factors, miners can only get revenue from published blocks. Based on this, we define \emph{absolute revenue} as the number of valid blocks obtained per unit of time. In Bitcoin system, we take 10 minutes as unit time. 
	
	Through the state machine, in the $i^{th}$ adjustment interval (eg. difficulty adjustment period), $n_i$ blocks will appear on the longest public chain and $m_i$ blocks are mined totally during one attack round~(eg. from stage(0,0,0) back to stage (0,0,0)). In addition, we use $T_i$ to represent the total time spented in the $i_{th}$ adjustment interval. Considering the change of computing power, the $S_i$ is used to represent the Hashrate of total system, the $math_i$ and the $t_i$ represent the theoretical time and the actual time that is spent mining one block during the $i^{th}$ adjustment interval respectively. Take Alice as an example, we can obtain the following equations:
	\vspace{-0.1cm}
	\begin{eqnarray}
	\label{eq:ni}
	n_i=R_1+R_2+R_h \quad \label{eq:mi}m_i=1\
	\vspace{-0.9cm}
	\end{eqnarray}
	
	Eq. \eqref{eq:ni} give us that it costs $T_1$ during the first difficulty adjustment period. 
	\vspace{-0.2cm}
	\begin{equation}
	T_1={2016*m_1*t_1}/{n_1}\quad math_1=1\;unit\; time\\
	\end{equation}
	\vspace{-0.5cm}
	\begin{equation}
	t_1=math_1*{S_0}/{S_1}\quad S_0=1;
	\end{equation}
	\vspace{-0.5cm}
	
	After the first period, the system will adjust the difficulty to satisfy mining one block per ten minutes. We can obtain the new average time of blocks generation during the $i_{th}$ period. Alice's \emph{absolute revenue} can be expressed as Eq. \eqref{eq:absolute}:
	\vspace{-0.2cm}
	\begin{equation}
	R_{absolute}=
	\label{eq:absolute}
	{\sum_{i=1}^k\frac{2016*R_1}{n_i}}/{\sum_{i=1}^kT_i} 
	\end{equation}
	\vspace{-0.3cm}
	\begin{equation}
	T_i={2016*m_i*t_i}/{n_i}\quad n_0=1
	\end{equation}
	\vspace{-0.5cm}
	\begin{equation}
	math_i=math_{i-1}*{2016}/{T_{i-1}}\quad
	t_0=math_0
	\end{equation}
	\vspace{-0.5cm}
	\begin{equation}
	t_i=math_i*S_{i-1}/S_{i}\quad math_0=math_1
	\end{equation}
	
	\section{simulation results}
	\begin{figure}[t]
		\begin{minipage}[t]{1\linewidth}
			\setlength\abovecaptionskip{0.5pt}
			\setlength\belowcaptionskip{-1pt}
			\centering
			\includegraphics[width=0.9\textwidth,height=0.18\textheight]{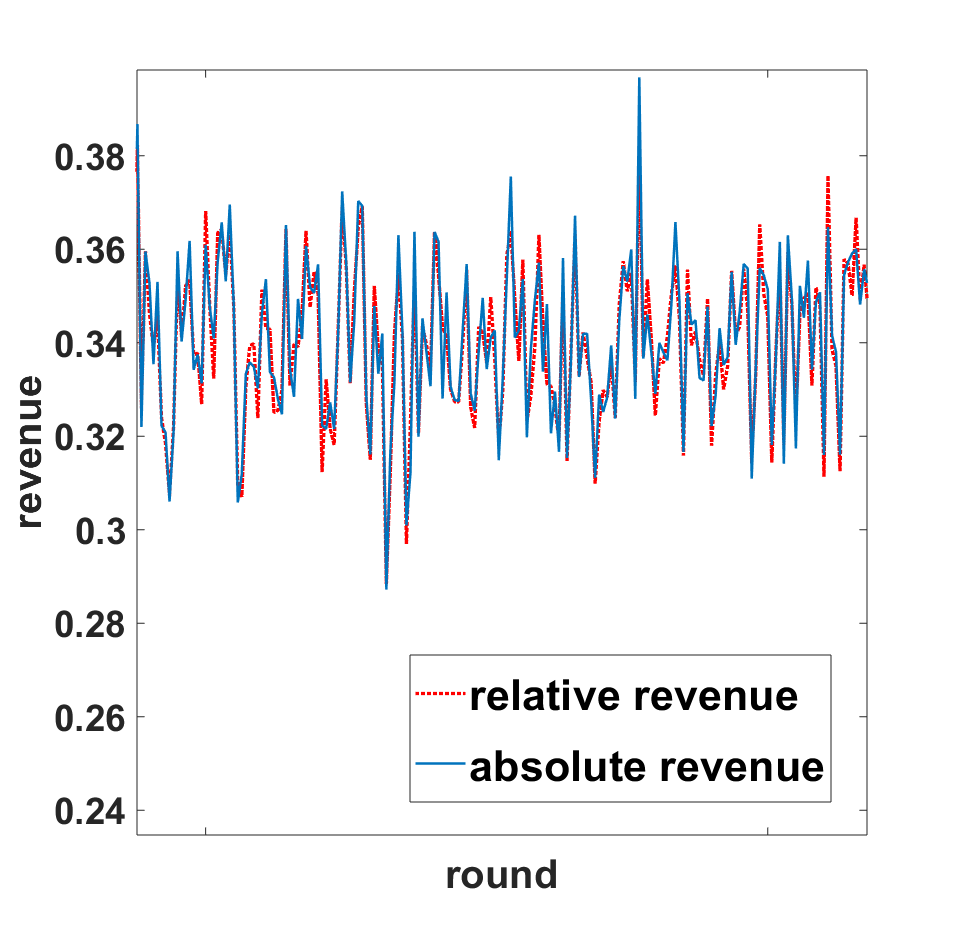}\\
			\caption{Relative revenue and absolute revenue.}
			\label{fig:relative}
		\end{minipage}
		\vspace{-0.6cm}
		\hspace{0.2cm}
		
	\end{figure}
	\begin{figure}[t]
		\begin{minipage}[t]{1\linewidth}
			\setlength\abovecaptionskip{-0.5pt}
			\setlength\belowcaptionskip{-1pt}
			\centering
			\includegraphics[width=0.8\textwidth,height=0.2\textheight]{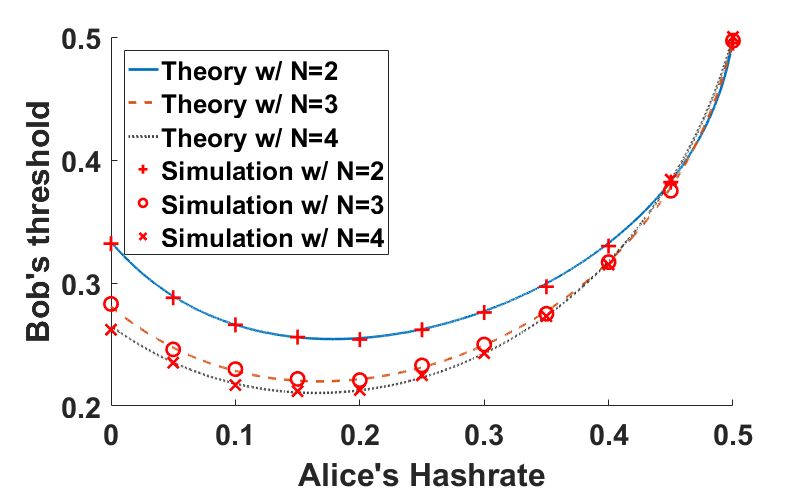}\\
			\caption{Bob's threshold under the influence of Alice’s Hashrate.}
			\label{pic:exp_basic_group2}
		\end{minipage}
		\hspace{0.1cm}
		\vspace{-0.6cm}
	\end{figure}
	
	\begin{figure*}[t]
		\begin{minipage}[t]{0.3\linewidth}
			\setlength\abovecaptionskip{-0.5pt}
			\setlength\belowcaptionskip{-1pt}
			\centering
			\includegraphics[width=0.9\textwidth,height=0.2\textheight]{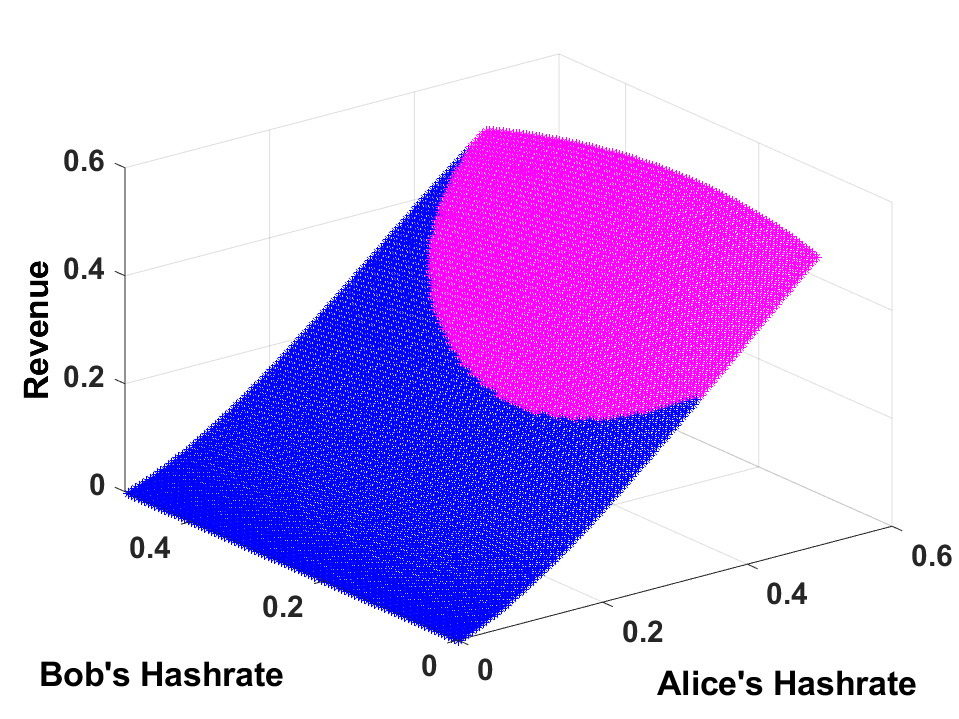}\\
			\caption{Alice's revenue w.r.t. Alice and Bob's Hashrate with $N{=}2$.}
			\label{pic:exp_basic_ber2}
		\end{minipage}
		\hspace{0.2cm}
		\begin{minipage}[t]{0.3\linewidth}
			\setlength\abovecaptionskip{-0.5pt}
			\setlength\belowcaptionskip{-1pt}
			\centering
			\includegraphics[width=0.9\textwidth,height=0.2\textheight]{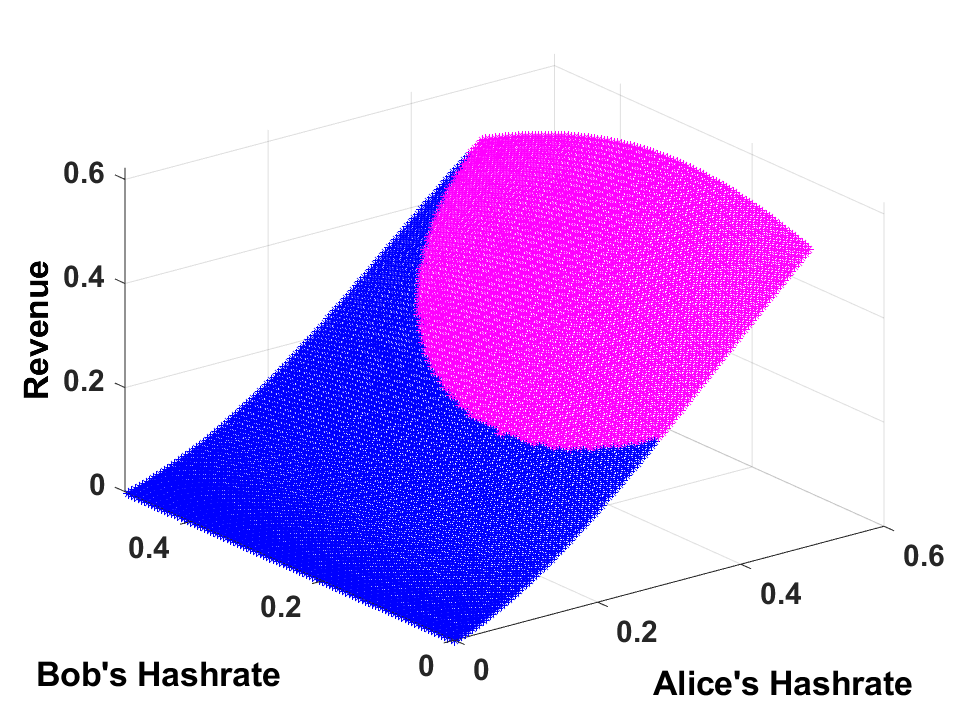}\\
			\caption{Alice's revenue w.r.t. Alice and Bob's Hashrate with $N{=}3$.}
			\label{pic:exp_basic_ber3}
		\end{minipage}
		\begin{minipage}[t]{0.3\linewidth}
			\setlength\abovecaptionskip{-0.5pt}
			\setlength\belowcaptionskip{-1pt}
			\centering
			\includegraphics[width=0.9\textwidth,height=0.2\textheight]{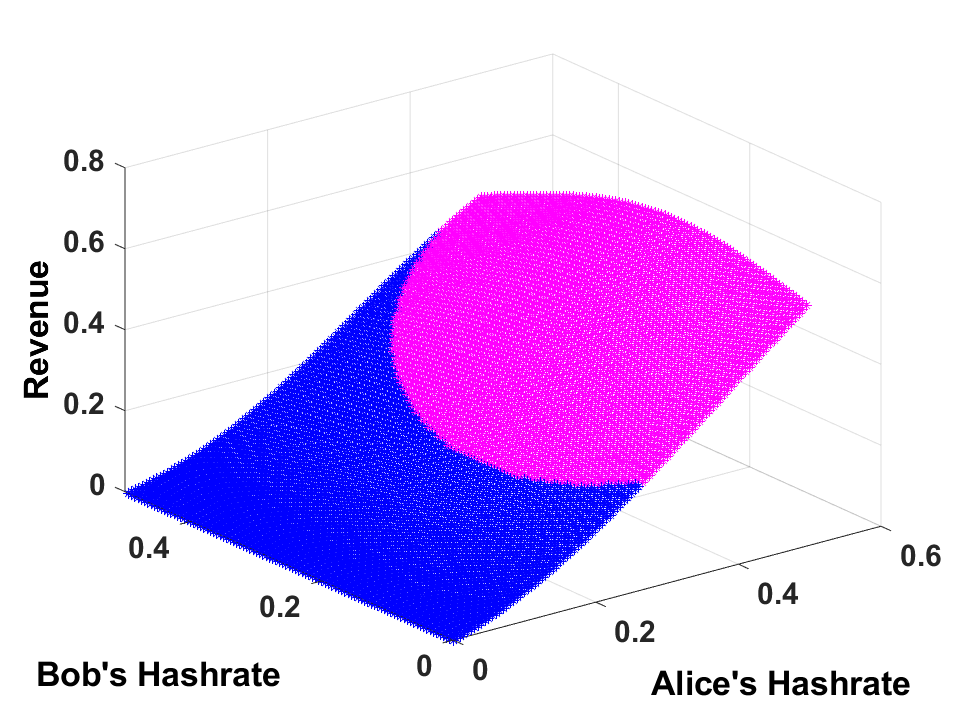}\\
			\caption{Alice's revenue w.r.t. Alice and Bob's Hashrate with $N{=}4$.}
			\label{pic:exp_basic_ber4}
		\end{minipage}
		\hspace{0.1cm}
		\vspace{-0.5cm}
	\end{figure*}
	
	\begin{figure*}[t]
		\begin{minipage}[t]{0.32\linewidth}
			\setlength\abovecaptionskip{3pt}
			\setlength\belowcaptionskip{-1pt}
			\centering
			\includegraphics[width=0.9\textwidth,height=0.15\textheight]{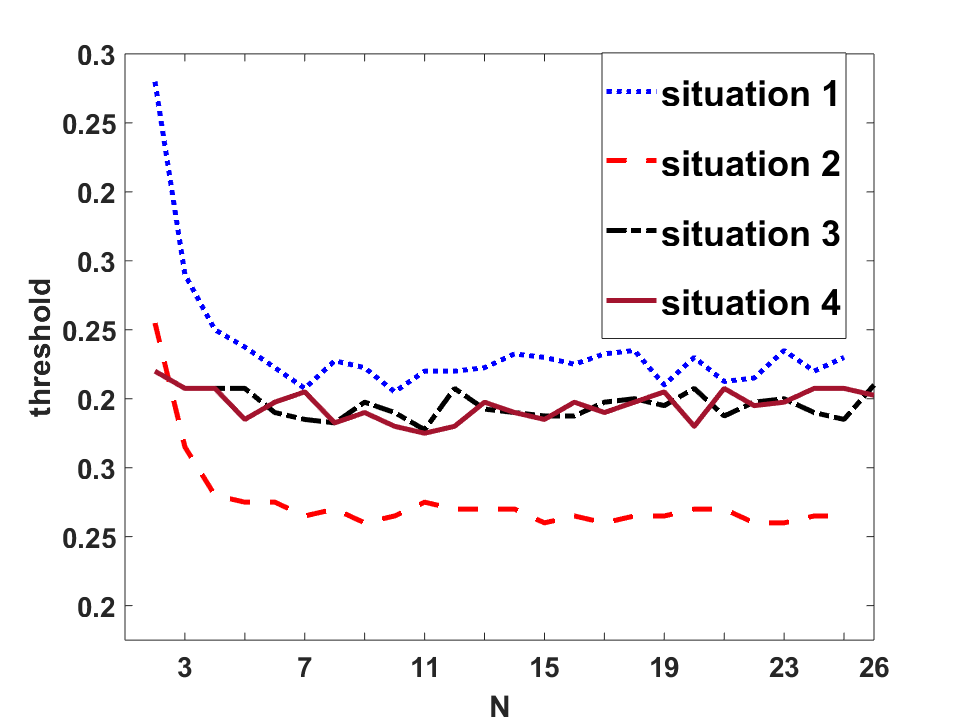}\\
			\caption{Threshold convergence process.}
			\label{fig:converg}
		\end{minipage}
		\begin{minipage}[t]{0.32\linewidth}
			\setlength\abovecaptionskip{3pt}
			\setlength\belowcaptionskip{-1pt}
			\centering
			\includegraphics[width=0.9\textwidth,height=0.15\textheight]{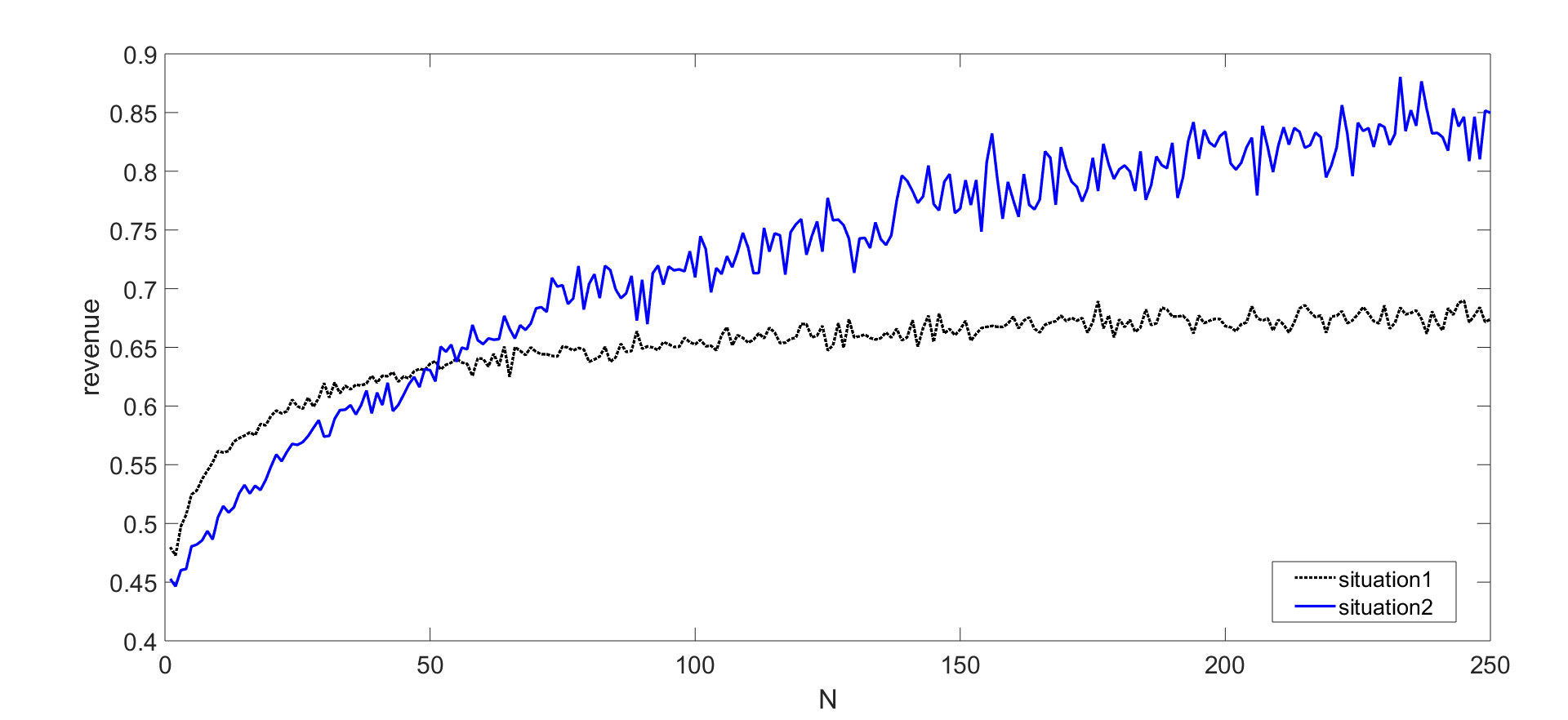}\\
			\caption{Upper limit.}
			\label{fig2}
		\end{minipage}
		\begin{minipage}[t]{0.32\linewidth}
			\setlength\abovecaptionskip{0.5pt}
			\setlength\belowcaptionskip{-1pt}
			\centering
			\includegraphics[width=0.9\textwidth,height=0.15\textheight]{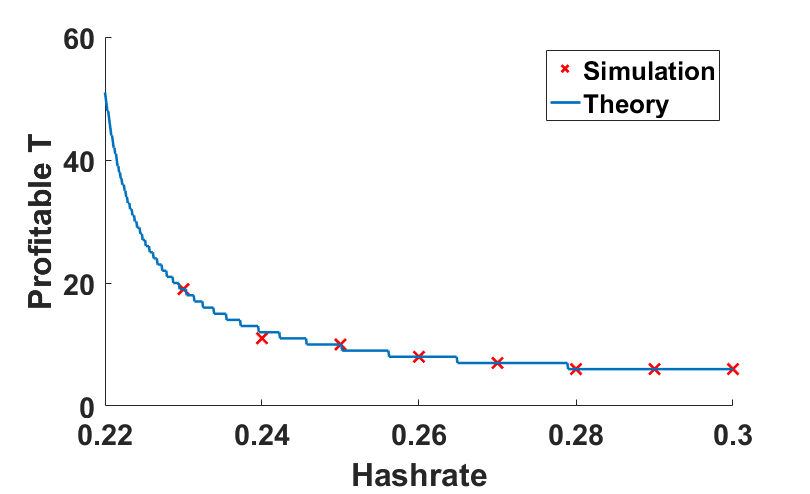}\\
			\caption{Profitable time and Hashrate.}
			\label{fig:round}
		\end{minipage}
		\hspace{0.2cm}
		\vspace{-0.7cm}
	\end{figure*}
	
	In order to verify the validity of theoretical analysis, we compare them with the results of a Bitcoin system simulator in this section. We set the block generation process to be exponentially distributed and run the simulator a million times. Based on the simulation results and theoretical results, we phrase the following observations:
	
	\begin{lemma} {\textit {When there are multiple attackers in Bitcoin system, the attackers' minimum profitable thresholds decrease and the system security is degraded.}}\end{lemma} 
	
	When there is only one attacker in system, \cite{ref:majority} proposed that when there are branches, if \begin{math}\gamma_1{=}\gamma_2{=}1/2\end{math}, the profitable threshold for attacker is 25\% . \cite{ref:ruanna} shows when there are two attackers with same Hashrate, the profit threshold will be lower than 25\% and it is easier to launch selfish-mining. We model this process with state machine shown in section \uppercase\expandafter{\romannumeral3} and the mathematical model verify this conclusion well.
	
	We consider the situation that $\gamma_{1}{=}\gamma_{2}{=}1/2$ and $\theta_{1}{=}\theta_{2}{=}1/3$. Driving Eq. \eqref{eq:RA}, we can obtain that when Alice's Hashrate is 16\%,  Bob's profitable threshold can reach the minimum: 21.06\%. When Alice's Hashrate is less than 16\%, the derivative of Eq. \eqref{eq:RA} is greater than 0, which means that Bob's threshold relative to Alice’s Hashrate is monotonically decreasing. When Alice's Hashrate is more than 16\%, the derivative of Eq. \eqref{eq:RA} is less than 0, which means that Bob’s threshold relative to Alice's Hashrate is monotonically increasing. In Figure \ref{pic:exp_basic_group2}, the blue curve represents the theoretical result and the red dots represent the simulation results. Three blue curves represent three cases: $N$ is 2, 3 and 4. We can observe that when Alice's Hashrate is around 16\%, Bob's threshold can be minimum. Through calculation and simulation, attackers' profitable threshold is 27\%, 23\% and 22\% when $N$ is 2, 3 and 4 respectively if Alice and Bob own the same Hashrate. It shows that when there are two attackers, they can adopt strategies to successfully attack with less than 25\% of total Hashrate.
	
	Figure \ref{fig:converg} also proves this result. The blue curve and the red curve represent that when there is only one attacker(we call it situation 1) and two attackers(we call it situation 2) in Bitcoin system, the relationship between $N$ and threshold. It shows that under same condition, the threshold of situation 1 is always higher than the threshold of situation 2.
	
	After \cite{ref:majority} published, people realized that the mining pool with more than 25\% of the Hashrate can successfully attack, so the system constrains the Hashrate of the mining pool to defend against the attack. We prove that this is not enough through the state machine model. In fact, it's much easier to attack than our current cognition. The Bitcoin system is easier to be attacked and its security is much more fragile.
	
	\begin{lemma}{\textit { If $N$ is no larger than 4, there is a negative correlation between Bob's lowest profitable threshold and $N$ while his revenue and $N$ are positively correlated. In the Bitcoin system, whether there is one attacker or two attackers, the profitable threshold will converge with the growth of $N$.}}\end{lemma}
	
	The lowest threshold is decreasing as N becomes larger.  We use Figure \ref{pic:exp_basic_ber2}, Figure \ref{pic:exp_basic_ber3} and Figure \ref{pic:exp_basic_ber4} to describe the revenue situations. Those three images represent Bob's revenue when two attackers' Hashrate are changing separately. Since we consider $\gamma_{1}=\gamma_{2}={1}/{2}$ and $\theta_{1}=\theta_{2}={1}/{3}$ in this current situation, Alice's and Bob's revenue are symmetrical. In these figures, blue part is the revenue and purple part highlights the moment Bob can gain additional income from the attack, in other words, the intersection of the blue and the purple parts is the threshold curve in Figure \ref{pic:exp_basic_group2}. 
	
	
	In Figure \ref{fig:converg}, situation 1 shows that when there is only one attacker, with the increase of $N$, the threshold convergences to 25\%. The convergence process tends to be smooth when attacker can own more than 5 private blocks. Situation 2 shows that when there are two attackers in system, the relationship between $N$ and threshold is consistent with one attacker, also a convergence process and its convergence speed is much faster. When $N$ is 4, it reaches the convergence balance, with threshold at 21.48\%. Situation 3 and situation 4 show that when Alice owns 25\% and 30\% Hashrate, Bob's threshold will also be a convergence process.
	
	That's because without destroying the normal operation of the system, Henry's Hashrate is at majority (this premise will be explained rationality in the next part). Based on this premise, in the real world situation, attackers can have small probability to own many private blocks and always take the leading position. 
	Hiding more private blocks can indeed increase attackers' revenue. However, a long private chain will easily expose the identity of the attacker, since a longer private chain can make it easier to distinguish it from normal blocks when they are published. On the other hand, without knowing the existence of another attacker, if $N$ is large, the risk to lose all it's private blocks gets higher. For the sake of insurance, the attacker might choose to disclose the number of private blocks to a certain extent to obtain corresponding income. In addition, this strategy can also rule out the impact of double-spending. Based on the above reasons, it is better to publish all private blocks once the length of private chain reaches 4, and start the next round of attack. \cite{ref:fruit} proposes that if we set up the timeliness of the block, we can effectively resist selfish mining attacks. The convergence of the threshold proves that this method is ineffective in current Bitcoin system, this is because in the current blockchain system, we default to a transaction requiring 6 valid blocks to be confirmed. Unfortunately, the threshold can reach convergence before six blocks.
	
	\begin{lemma}{\textit {In order to ensure the attack can proceed normally, \begin{math}\alpha_h >\max\{\alpha_1, \alpha_2\}\end{math} must be satisfied}}\end{lemma}
	
	As a counterexample, if Alice has the highest Hashrate, and there is no limit to the length of private chain, Alice can hide her private chain as long as possible. She can stay in the lead in most cases during the attack, which will lead to Alice's private chain becoming the only valid chain. In this case, Bob and Henry will choose to stop mining to reduce losses. We can speculate that under this circumstance, Alice's revenue can be close to 100\% and her attack actually becomes meaningless. This kind of attack is similar to 51\% attack. Simulation results also prove this. In Figure \ref{fig2}, when Alice's Hashrate is 45\%, Bob's Hashrate is 25\% and Henry's Hashrate is 35\%, we obtained situation 1. When Alice's Hashrate is 45\%, Bob's Hashrate is 35\%, and Henry's Hashrate is 25\%, we obtained situation 2. It shows that the longer attacker's private chain is, the more he can gain. As long as one attacker has the highest Hashrate, this situation could happen, regardless of how many Hashrate other miners have. According to this analysis, it is very meaningful to stipulate that Henry should have the highest Hashrate in the attack model.

	\begin{lemma} {\textit {The attackers will fail during the first difficulty adjustment period regardless of the attackers' Hashrate. However, he might gain profit after several periods, which is related to the attackers' Hashrate.}}
	\end{lemma}
	
	Assuming two attackers have the same Hashrate, we simulated and obtained Figure \ref{fig:relative}. \emph{relative revenue} and \emph{absolute revenue} are equal within the allowable range of error. Therefore, we can believe that the \emph{relative revenue} and \emph{absolute revenue} play the same role in representing benefit.
	
	As Eq. \eqref{eq:absolute} shows, when Alice has more Hashrate, she can get illegal revenue earlier. Figure \ref{fig:round} shows the simulation results match well with the theoretical result. The horizontal axis represents the attack round and the ordinate represents the attackers' revenue, also the blue curve is theoretical result and the red dots are simulation results. It shows that when attackers' Hashrate is relatively small, it takes a rather long period to gain profit. That means in the real system, it is a little bit hard to perform attack. If the global Hashrate increase, we can also use this formula to calculate when to stop the attack before we can benefit the most.
	
	\section{Conclusion}
	In this paper, we study how the existence of multiple misbehaving pools influences the profitability of selfish mining. By establishing the Markov chain model to describle the action of attackers and honest miners, we can obtain the minimum profitable threshold is symmetric 21.48\%. Considering the difficulty adjustment, we model the transient process and discover the negative correlation between the profitable time and the attackers' mining power.


\begin{thebibliography}{1}
		\bibitem{ref:whitepaper}
		S. Nakamoto. `` Bitcoin: A peer-to-peer electronic cash system'' , 2008.
		\bibitem{ref:majority}
		I. Eyal and E. G. Sirer. ``Majority is not enough: Bitcoin mining is vulnerable''. In \emph{Financial Cryptography and Data Security}. Springer, 2014, pp. 436-454.
		\bibitem{ref:ruanna}
		Q.H. Liu, N. Ruan, et al. ``On the Strategy and Behavior of Bitcoin Mining with N-attackers''. \emph{Proc. of the Asia Conference on Computer and Communications Security}, pp. 357-368, 2018.
		\bibitem{ref:data}
		https://bitinfocharts.com/comparison/bitcoin-hashrate.html
		\bibitem{ref:state}
		A. Papoulis, S. U. Pillai. Probability, random variables, and stochastic processes[M]. Tata McGraw-Hill Education, 2002. 
		\bibitem{ref:appendix}
		Technical report. http://medianet.azurewebsites.net/new-page/
		\bibitem{ref:fruit}
		R. Pass, E. Shi ``Fruitchains: A fair blockchain''. \emph{Proc. of the Asia Conference on Computer Symposium on Principles of Distributed Computing}, pp. 315-324, 2017.
		\bibitem{ref:MDP}
		A. Sapirshtein, Y. Sompolinsky, A. Zohar. ``Optimal selfish mining strategies in bitcoin''. \emph{International Conference on Financial Cryptography and Data Security}, pp. 515-532, 2016.
	\end{thebibliography}
\end{document}